\documentclass[%
superscriptaddress,
 amsmath,amssymb,
 aps, physrev,
]{revtex4-2}

\usepackage{graphicx}% Include figure files
\usepackage{dcolumn}% Align table columns on decimal point
\usepackage{bm}% bold math
\usepackage[english]{babel}
\usepackage{float} 
\usepackage{xcolor}
\begin{document}

\title{\textbf{Coupled gas and bubble dynamics at the solidification front}}% 

\author{Bastien Isabella}
\affiliation{%
Université Lyon 1, CNRS\\
Institut Lumière Matière\\
UMR5306, F-69100, Villeurbanne, France
}%

\author{Cécile Monteux}
\affiliation{%
 Sciences et Ingénierie de La Matière Molle\\
 UMR 7615, ESPCI Paris, PSL Research University, CNRS, Sorbonne Universités\\
 75005 Paris, France
}%

\author{Sylvain Deville}
\email{Corresponding author: sylvain.deville@univ-lyon1.fr}
\affiliation{%
Université Lyon 1, CNRS\\
Institut Lumière Matière\\
UMR5306, F-69100, Villeurbanne, France
}%

\date{\today}

\begin{abstract}
The formation and entrapment of gas bubbles during solidification significantly influence the microstructure and mechanical properties of materials, from metallic alloys to ice. While gas segregation at the solidification front is well-documented, the real-time dynamics of bubble nucleation, growth, and engulfment-and their dependence on solidification velocity-remain poorly understood. In this study, we use in situ cryo-confocal fluorescence microscopy to investigate the coupled gas-bubble dynamics at the solidification front of carbonated water, systematically varying the solidification velocity ($V = 1-20 \mu m/s$) while maintaining a constant thermal gradient ($G = 15 K/mm$). Our experiments reveal that bubble nucleation is governed by a characteristic nucleation time, which emerges from the interplay between gas diffusion ahead of the front, nucleation kinetics, and bubble growth, all competing with the advancing solidification front. These results allow us to estimate the critical gas concentration for bubbles nucleation in carbonated water. These results offer a detailed understanding of the mechanisms controlling bubble nucleation and entrapment during solidification at constant thermal gradient. They contribute to the development of strategies to control bubble formation in industrial processes where the presence of bubbles can either be detrimental or intentionally harnessed.
\end{abstract}

%Use showkeys class option if keyword
               %display desired
\maketitle

\section{Introduction\label{sec:introduction}}

The occurrence of bubbles during solidification processes is of considerable significance across various disciplines, including environmental sciences with studies on ice cores~\cite{ahn2008co2}, lake ice~\cite{gow1977growth}, and hailstones~\cite{knight1968final}, as well as in engineering domains such as metallurgy~\cite{zhang2013nucleation}, crystal growth~\cite{li2013bubbles, bouaita2019seed}, materials processing~\cite{liu2022bubble}, and cryopreservation~\cite{korber1988phenomena}. Bubbles can contribute to the development of porosity in solidified materials, thus affecting their mechanical and functional properties by altering their microstructure~\cite{ghezal2012observation}. While porosity is typically regarded as a defect in solid materials, a comprehensive understanding of the mechanisms and factors that lead to bubble entrapment by the solidification front might facilitate its control, thus enhancing the properties of the resulting materials. Conversely, in applications where porous materials are desired, an in-depth understanding of these mechanisms could enable the engineering of materials with precisely controlled porosity characteristics, including the number, size, and morphology of trapped bubbles~\cite{liu2022bubble}. Parameters such as the composition of the liquid, its temperature, the temperature gradient within the system, and the solidification velocity are potential key factors influencing bubble entrapment. 

During solidification objects residing within the liquid phase may either be pushed or engulfed by the solidification front~\cite{tyagi2020objects, asthana1993engulfment, stefanescu1988behavior}. The nature of their interaction with the solidification front is significantly influenced by the presence of solutes and gases, particularly in proximity to the solidification front~\cite{tyagi2022solute}. Indeed, during directional solidification, solutes have a tendency to segregate at the solidification front, thereby establishing a concentration gradient between the bulk liquid and the solidification front~\cite{tiller1953redistribution, yoshimura2008growth}. This solute segregation at the interface results from their limited solubility in the solid phase~\cite{kurz2023fundamentals}, and the effect is further exacerbated by the proximity of neighboring objects, which perturb the local diffusion gradients.

The mathematical model formulated by Pohl~\cite{pohl1954solute} enables the prediction of solute and gas segregation profiles at the solidification front as a function of factors such as solidification velocity, solute diffusivity, and solubility within the liquid phase, assuming negligible diffusion within the solid phase, an absence of vaporization, and purely diffusive transport within the liquid. Assuming the absence of convection in the liquid phase and negligible gas diffusion in the solid, the variation in gas concentration during solidification can be estimated using the following equation :

\begin{equation}
  \frac{C_{L}(x, t)}{C_{0}}=1+\left(\frac{1-k}{k}\right)\left[\exp \left(-\frac{V_{s f}}{D} x\right)-\exp \left(-\frac{V_{s f}(1-k)\left(x+k V_{s f} t\right)}{D}\right)\right]
  \label{eq:concentration_profile}
\end{equation}

where $V_{sf}$ is the solidification front velocity ($m/s$), $t$ is the time ($s$), $D$ is the diffusion coefficient of the gas ($m^{2}/s$), $C_L$ is the gas concentration at a distance $x$ from the solidification front and $C_0$ is the bulk gas concentration in the liquid far from the solidification front ($g/L$). The partition coefficient $k$ is related to the solubility of the gas in the ice and is defined as the ratio of gas solubility in the solid phase to the one in the liquid phase (and is therefore dimensionless). The local segregation of solutes is especially critical for examining the morphology of the solidification front, the occurrence of constitutional supercooling leading to the formation of premelted liquid films~\cite{wettlaufer1999impurity, dedovets2018five}, as well as for understanding the behavior of objects in proximity to the front~\cite{tyagi2022solute}.

The continued segregation of gas at the solidification front can therefore lead to a marked increase in gas concentration, potentially exceeding the solubility limit within the liquid medium and eventually initiate bubble nucleation. Classical nucleation theory~\cite{becker1935kinetische,volmer1926keimbildung, frenkel1946kinetic} provides a theoretical framework for characterizing the emergence of new phases, such as gas bubbles in supersaturated liquids. This theory postulates that random thermal fluctuations may exceed the local free energy barrier required for nucleation~\cite{ward1970thermodynamics}, consequently allowing for the formation of bubbles.~\cite{bowers1995supersaturation}. The classical nucleation theory enables the estimation of the likelihood that a nucleus will evolve into a stable bubble~\cite{cavitation,baidakov2014spontaneous}, the theoretical nucleation rate~\cite{kalikmanov2012classical,blander1975bubble}, and the critical nucleation concentration~\cite{bowers1995supersaturation} within a given system.

Throughout the process of solidification, the nucleation free energy barrier is primarily influenced by surface tension and pressure differences, both of which depend on the temperature of the liquid and the concentration of gas. Although the mechanisms responsible for bubble nucleation are well understood, the dynamics associated with bubble nucleation during solidification are not well elucidated. Bubble nucleation may occur homogeneously, within the bulk of the liquid in regions of gas enrichment, or heterogeneously, on a substrate such as the front of solidification itself~\cite{jones1999bubble}. Consequently, determining the critical nucleation concentration poses a challenge due to the significant variability in values derived from distinct experimental measurements or theoretical predictions~\cite{jones1999bubble, geguzin1981crystallization, lipp1987gases}.

The dynamics of bubbles during solidification represent a complex process encompassing multiple mechanisms. \textit{In situ} experimental investigations of bubble growth~\cite{Bari_Hallett_1974, meijer2024bubble, Werner2021-gc, Cao2022intermetallic, lipp1987gases} and interactions with the solidification front~\cite{yoshimura2008growth, Vasconcellos1975-qo, Murakami2002-al, Maeno1967-ck, ZHDANOV1980659, Thievenaz2025bubble,Wei2000-ou, WEI201779} have been largely examined. Conversely, the study of bubble nucleation dynamics, an essential phase of bubble dynamics during solidification, remains underexplored. The existing understanding of nucleation during solidification primarily stems from theoretical modeling or post-solidification analysis, and inquiries into nucleation dynamics are notably scarce in the literature. In numerous studies, experimental conditions during solidification, such as solidification velocity and thermal gradient, were not consistently controlled, often varying throughout the process. Although these approaches afford some understanding of nucleation phenomena, they are insufficient for probing nucleation dynamics under strictly defined and constant conditions~\cite{yoshimura2008growth,carte1961air,Murakami2002-al,Thievenaz2025bubble}.

Here we examine \textit{in situ} the dynamics of bubble nucleation during the directional solidification of carbonated water using cryo-confocal fluorescence microscopy. This method provides precise control over both the solidification velocity and the temperature gradient throughout the solidification process. A comprehensive examination was conducted focusing on the number and location of bubble nucleation events, their growth and size, and their subsequent engulfment by the solidification front. The analysis was performed under carefully controlled solidification velocities and temperature gradients to evaluate the influence of these parameters on the coupled dynamics of these processes.

\section{Materials \& methods\label{sec:materials}}

\subsection{Materials and sample preparation\label{ssec:sample_preparation}}

An aqueous fluorophore, Sulforhodamine B, was obtained from FluoTechnik, alongside a Sodastream Terra device to facilitate the injection of carbon dioxide into water. The solutions were prepared by dissolving  Sulforhodamine B in distilled water, followed by stirring for approximately one minute, to obtain a $10^{-5}M$ solution. Carbon dioxide was introduced into the solution by administering three one-second pulses (controlled with a timer) using the Sodastream Terra device. The pH was measured using a calibrated electronic pH meter, and the carbonate hardness (KH) was determined using commercially available KH test strips. Both measurements were used to determined the initial $CO_2$ concentration before freezing. The temperature (room temperature) was kept constant for all experiments. Subsequently, the solution was immediately introduced into a rectangular Hele-Shaw cell via capillary action. This cell comprised two glass slides (Menzel, thickness $0,13-0,19mm$), held apart by two spacers composed of double-sided adhesive tape, which maintained a consistent sample thickness of approximately $100\mu m$. The cell was sealed on one side with a plastic support (used to pull the sample through the temperature gradient) and on the opposite side with nail polish to prevent both evaporation and leakage. The reproducibility of the approach was assessed by repeating the sample preparation and running the same experiment several times.

The use of thin Hele-Shaw cell is standard in solidification studies, their benefits and relevance have been reported multiple times, for example in the context of solidification studies of transparent alloys.

\subsection{Experimental setup and typical solidification parameters\label{ssec:setup}}

\begin{figure}[!htbp]
  \centering
  \includegraphics[width=8cm]{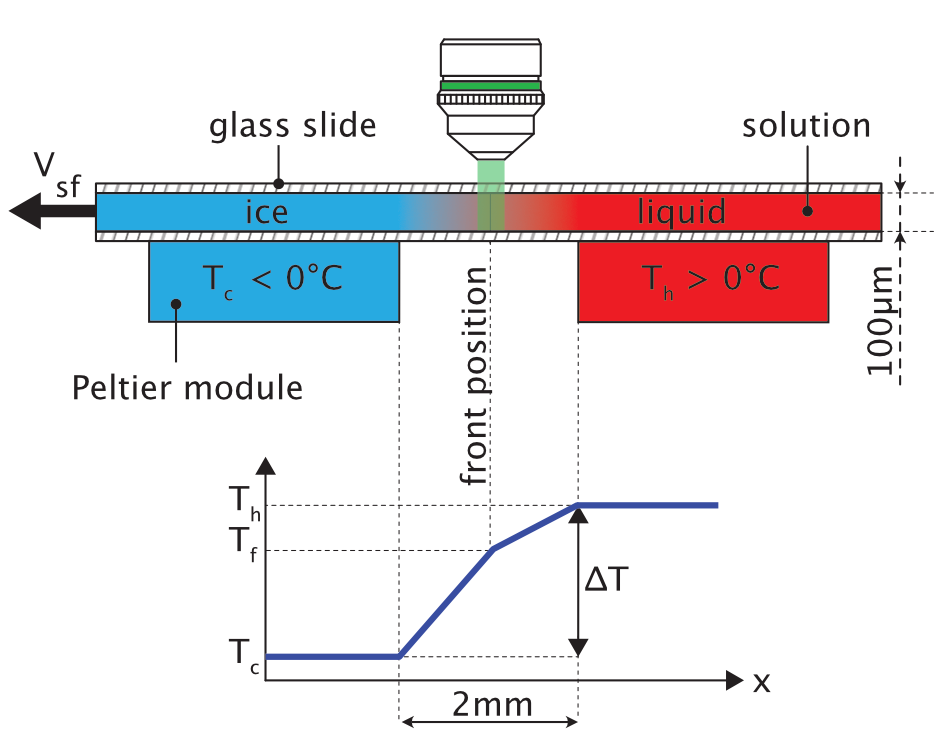}
  \caption{Experimental setup for cryo-confocal microscopy. A constant temperature gradient $\Delta T = T_h - T_c$ is created between the Peltier modules, specifically within the gap ($d=2mm $). The sample is translated through the temperature gradient at a constant velocity $V_{sf}$, thereby inducing ice crystal growth at a velocity $V_{sf}$. Consequently, the interface remains at a fixed position within the observation frame. The samples possess a thickness of $100\mu m$, and $T_f$ denotes the liquid's freezing point. Due to differing thermal conductivities, the temperature gradient in the ice is greater than in the liquid. Within the study, the reported magnitude of the temperature gradient $\Delta T$ corresponds to the temperature difference between the two Peltier modules.}
  \label{fig:figure1}
\end{figure}

Unidirectional solidification experiments were conducted utilizing the experimental setup described previously~\cite{dedovets2018five} and visualized through confocal fluorescence microscopy (Fig.~\ref{fig:figure1}). The Hele-Shaw cell was positioned beneath the microscope objective on two Peltier modules and was pulled through the temperature gradient at a constant velocity by a stepper motor (Micos Pollux Drive stepper motor with VT-80 translation stage (PI, USA)). The temperature gradient was established by two Peltier modules. The temperature is measured through thermocouples placed at the surface of the Peltier and continuously monitored. The temperature gradient is managed with a TEC-1122 Dual Thermo Electric Cooling Temperature Controller from Meerstetter Engineering, Switzerland. This setup enables to control independently the temperature gradient $\Delta T = T_h - T_c$ created between the Peltier modules and the growth velocity  $V_{sf}$ of the freezing front. This ability is critical since, as demonstrated in this paper and in previous studies, many of the phenomena associated to solidification are very sensitive to both parameters. We performed all experiments with $T_h = 15 ^\circ C$ and $T_c = - 15 ^\circ C$.

The stage was enhanced by enclosing it within a Plexiglas box supplied with dry air to minimize humidity (typically between 5 and 10\%) to avert condensation and freezing on cold surfaces. This measure prevents potential friction with the sample and mitigates light scattering caused by frost.

The control of solidification velocity and temperature gradient is achieved independently, enabling the preservation of the solidification front at a stationary position beneath the microscope objective. Freezing experiment were conducted with velocities spanning from $1$ to $20\mu m/s$, alongside a temperature gradient of $15^{\circ}C/mm$.

In a typical experiment, the temperature gradient is kept constant, and the translation stage imposing the growth velocity applies a constant growth rate. The solidification front therefore reaches a steady state as soon as its position within the observation frame is constant. This typically occurs within seconds. All the data reported in the paper was obtained in a steady state regime, where the front velocity $V_{sf}$ is constant.

\subsection{Imaging\label{ssec:imaging}}

Images were obtained using a Leica TCS SP8 confocal laser scanning microscope (Leica Microsystems SAS, Germany), equipped with a long working distance (2.2mm), non-immersive objective (Leica HC PL APO 10x/0.40 CS2), to ensure an adequately large field of view and to minimize the influence of the microscope's thermal mass on the freezing process. A 488nm laser was employed for all experiments. Image acquisition was conducted in resonant mode at a resolution of $512\times512$ pixels ($582 \times 582\mu m$), resulting in acquisition times ranging from 0.133s to 0.333s per frame. A single photo-detector was utilized, operating between 500nm and 795nm, which corresponds to the emission range of Sulforhodamine B when dissolved in water. Air bubbles do not fluoresce and thus appear black in the images, akin to ice, which also appears black due to the negligible solubility of fluorophores in ice. Previous experiments have demonstrated that Sulforhodamine B does not exhibit any significant impact on the freezing behavior of the system.

\section{Results and discussion\label{sec:results_and_discussion}}

\subsection{Experimental observations\label{ssec:experimental_observations}}

\begin{figure*}[!htbp]
  \centering
  \includegraphics[width=16cm]{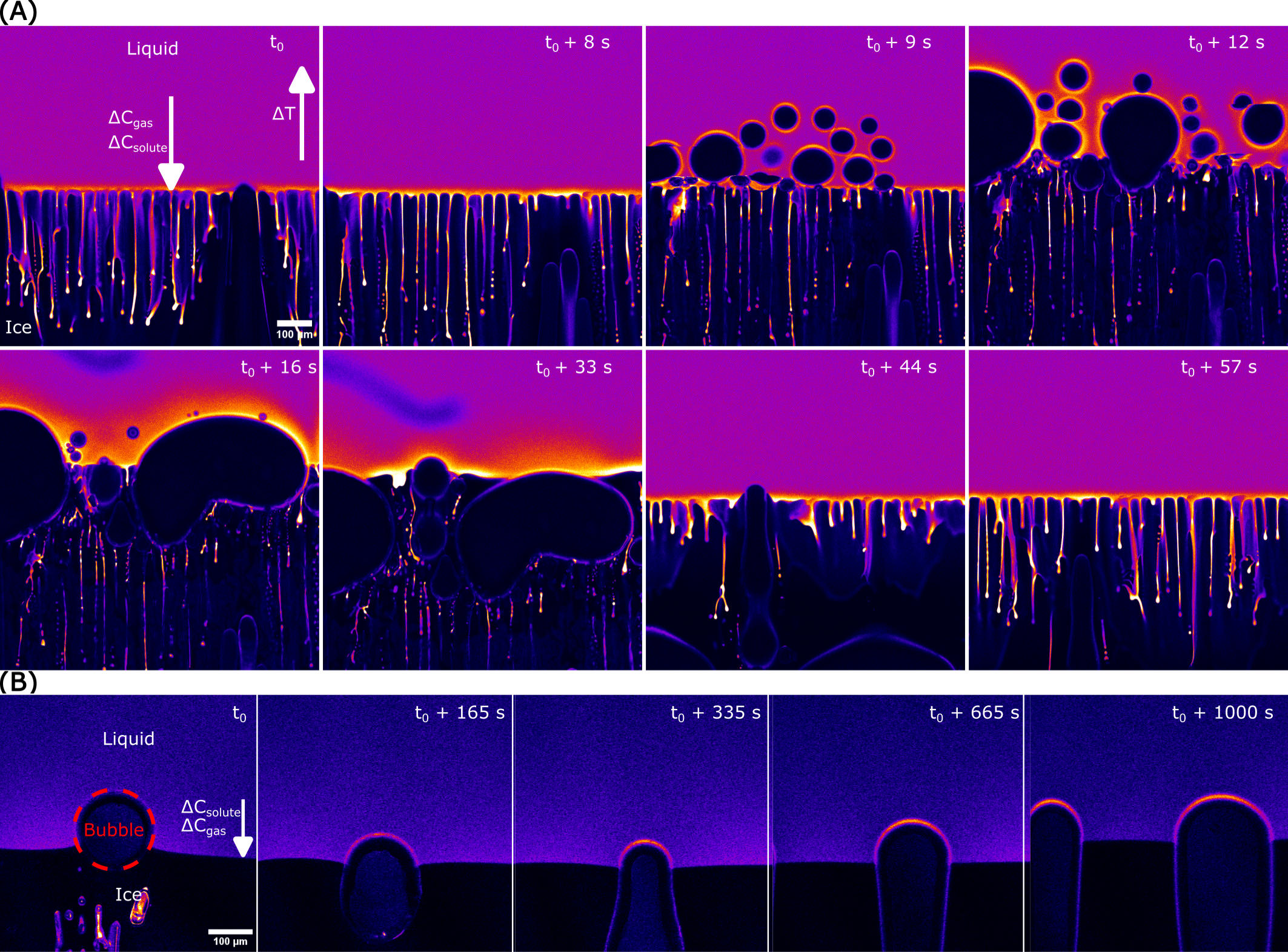}
  \caption{Representative images obtained through confocal cryomicroscopy. These time-lapse sequences of typical two-dimensional confocal images illustrate the freezing process of carbonated water and the different stages encountered during a typical solidification experiment. In each image, the lower, darker region corresponds to the ice phase, whereas the upper, lighter region corresponds to the liquid phase. Bubbles, being non-fluorescent, appear as dark regions within the liquid. (A) The time-lapse sequences display typical 2D cryo-confocal images demonstrating the nucleation, interaction, and engulfment of bubbles during the solidification process. Velocity of the solidification front $V_{sf} = 20 \mu m/s$. The bubble formation in these images is ascribed to nucleation induced by gas accumulation at the advancing solidification front, attributed to gas segregation. (B) The sequences also illustrate the formation and evolution of cylindrical bubbles at the solidification front during solidification for  $V_{sf} = 1 \mu m/s$. For each panel, $t_0$ corresponds to the time of the first frame of the sequence shown in the panel.}
  \label{fig:figure2}
\end{figure*}

We conducted horizontal solidification experiments by displacing a Hele-Shaw cell filled with carbonated water, at varying freezing velocities $V_{sf}$. A representative series of time-lapse 2D confocal images illustrating the freezing process is presented in Fig.~\ref{fig:figure2}A for $V = 20\mu m/s$. 

In the observational frame, the solidification front remains stationary, whereas in the sample's reference frame, the solidification front progresses through the medium at a velocity $V_{sf}$. The ice appears black and the aqueous solution pink. Bubbles nucleate near or at the solidification front within several seconds of freezing. This process is attributed to the segregation of gas at the front, which results in a localized increase in gas concentration surpassing the critical nucleation threshold under the given experimental conditions. 

We performed preliminary observations by imaging the sample at various depths during solidification, and in particular close to the glass slides, to assess the influence of the glass slide on the bubble nucleation behavior. The glass slide do not appear to be preferential nucleation sites for the bubbles, and bubbles do not appear to stick to the glass slide. However, it is still possible that bubble nucleation occurs onto the glass slides and that bubble nearly instantaneously detach from the glass during the initial burst growth stage. All subsequent experiments were therefore performed in the middle of the sample.

The following images, from $t=9s$ to $t=12s$, demonstrate that bubbles can experience an increase in size as a consequence of gas segregation and diffusion or through the coalescence of adjacent bubbles. 

The transfer of gas between a bubble and the surrounding liquid is regulated by the Laplace pressure, which is a function of the bubble's curvature. The bubbles eventually interact with the solidification front and undergo deformation during their engulfment ($t=16s$ and $t=33s$). Mechanical stresses acting on bubbles arise from the pushing effect of the solidification front. This mechanism has been described for solid particles ~\cite{Rempel2001}, and bubbles experience a similar effect~\cite{Park2006}. After engulfment of the bubbles by the front ($t=44s$), another cycle of gas segregation is initiated ($t=57s$), until the next nucleation event.

The deformation experienced by a bubble during its interaction with the front is dictated by the equilibrium between gas segregation and diffusion at the solidification front, as well as the solidification velocity $V_{sf}$. This equilibrium can result in diverse morphologies of bubbles entrapped within the solid; some bubbles retain their spherical shape, some others undergo minor deformation resulting in egg-like shapes, and some are stretched during the interaction to become cylindrical. 

The deformation of these cylindrical bubbles and the influence of solidification velocity, gas concentration, pressure, and mass transport of gas have been thoroughly examined in previous studies~\cite{liu2022bubble,Bari_Hallett_1974,Maeno1967-ck}. Moreover, models have been formulated to predict the formation of such bubbles~\cite{Wei2000-ou,Thievenaz2025bubble}. Although a sharp transition from circular to cylindrical bubbles was predicted~\cite{Thievenaz2025bubble}, we did not observe it here. In our steady state regime, with a constant front velocity and constant temperature gradient, there is a continuous transition from circular to cylindrical bubbles. The stability, and thus the time that cylindrical bubbles remains at the front, increases as the front velocity decreases. Cylindrical bubbles persist longer (up to $1000s$ and more) at reduced solidification velocities ($1-2\mu m/s$, with some observed to grow laterally when the solidification velocity $V_{sf}$ is below $5\mu m/s$ (Fig.~\ref{fig:figure2}C). Conversely, at elevated solidification velocities $V_{sf}$ ($5-20\mu m/s$), cylindrical bubbles may still form; however, they are smaller and become engulfed by the solidification front shortly after formation (Fig.~\ref{fig:figure2}B). This is attributed to the inadequate gas diffusion from the surrounding liquid to the exposed tip of the cylindrical bubbles at high solidification velocities. 

Since the nucleation of bubbles occurs close to the solidification front, this also means there is a limited amount of time (typically a few seconds) during which the bubbles could interact with the glass slides, before being engulfed by the solidification front. Although $CO_2$ bubbles are known to wet glass, it seems likely that the rapid kinetics observed here prevent such wetting from occurring. We therefore believe that the glass slides have little influence on the observations reported in this study.

Our focus now shifts to the dynamics of bubble nucleation.

\subsection{Average nucleation rates \label{ssec:nucleation_rates}}

\begin{figure}
  \centering
  \includegraphics[width=8cm]{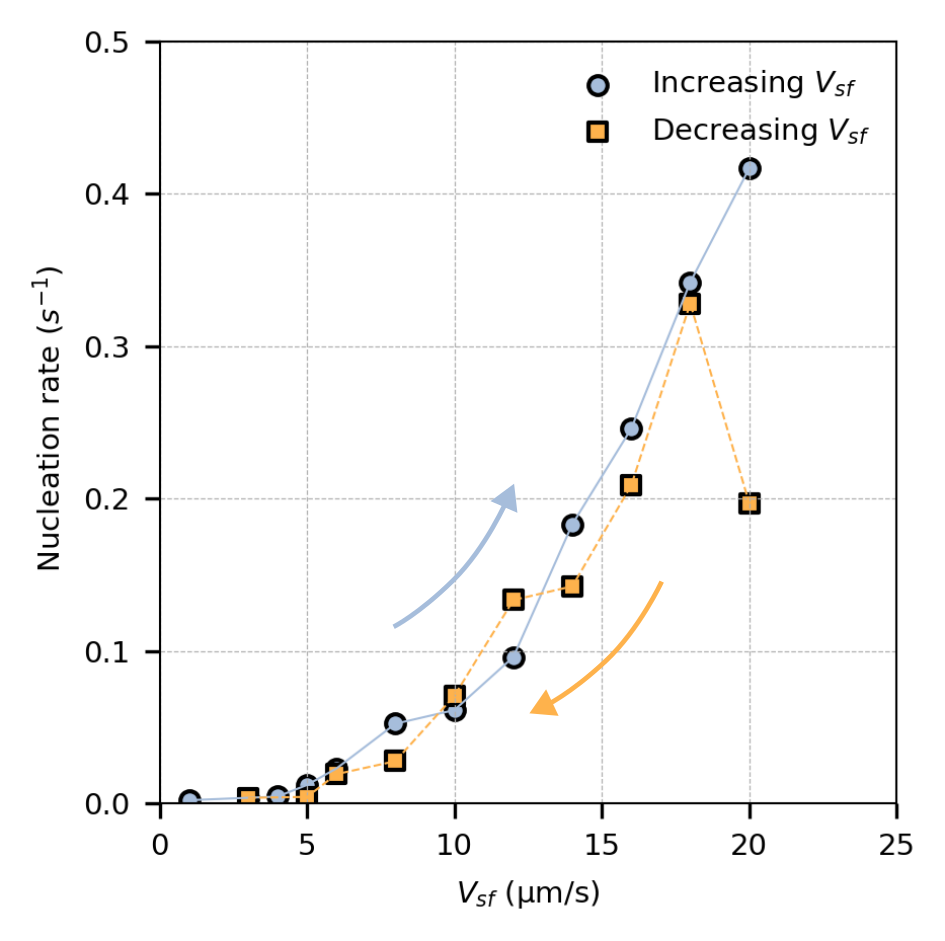}
  \caption{Evolution of the nucleation rate of bubbles as a function of the velocity of solidification $V_{sf}$. Measurement were done with similar sample and identical experimental conditions but with either an increasing or a decreasing solidification front velocity. For each data point in this figure, the motor velocity is set to a specific prescribed value. At this constant motor velocity, the number of observed bubble nucleation events is recorded in order to determine the mean nucleation rate corresponding to that condition. Following each measurement, the motor velocity is adjusted to a new value, and the procedure is repeated to obtain data at different velocity for a unique sample. For each velocity of solidification considered, the measurement time ranged from 3 minutes at the highest velocities of solidification to 9 minutes at the lowest velocities of solidification.}
  \label{fig:figure3}
\end{figure}

We first examine the impact of solidification velocity $V_{sf}$ on the average rate of bubble nucleation, quantitatively expressed as the number of bubbles formed per second (Fig.~\ref{fig:figure3}). This analysis currently focuses for now solely on the incidence of bubble nucleation events, deliberately excluding any subsequent phenomena such as changes in bubble size, growth kinetics, or interactions with adjacent bubbles. The freezing front velocity $V_{sf}$ is systematically increased from $1\mu m/s$ to $20\mu m/s$ (ramp up). Additionally, an alternative set of measurements is conducted by reducing the velocity from $20\mu m/s$ to $1\mu m/s$ (ramp down). 

The ramp up and ramp down procedures are the same. The velocity is held constant at a given value to complete the imaging work and then adjusted to a lower (ramp down) or higher (ramp up) value. When the velocity is changed to a new value, it takes only a few seconds for the front to stabilize around its new steady state position, at which point we start imaging. The solidification process continues long enough to collect sufficient statistics, typically between 20 and 70 for velocities in the range of $8-20 \mu m/s$ . At lower velocities, the number of bubbles remained very limited—typically fewer than 10 because of the small number of bubble nucleation. We keep track of the number of bubbles that appear during solidification and thus do not need to define a critical bubble size for nucleation and growth. The liquid being saturated with gas, we systematically observe that the bubble first go through an explosive growth stage. Considering the fast imaging rate used (up to 7.5 frames/s), we are thus able to determine precisely the moment when bubbles appear in the liquid. By using a single sample with a gradually increasing or decreasing solidification velocity, it is ensured that any observed variations in nucleation behavior are attributed to velocity changes rather than fluctuations in the initial gas concentration. 
 
We observe that the average nucleation rate increases with the solidification velocity for both the increasing or decreasing ramps. The enhanced nucleation rate observed at higher solidification velocities is consistent with an increased segregation of dissolved gases at the solidification front. Faster solidification front velocities accelerate local gas enrichment, making it more likely for the critical nucleation concentration $C_n^*$ to be reached within a shorter timescale.

Furthermore, nucleation rates demonstrate consistency and reproducibility across both velocity ramp protocols. A notable discrepancy is observed for $V = 20\mu m/s$, wherein the nucleation rate observed during the decreasing velocity ramp is markedly lower than that in the increasing velocity ramp. This inconsistency is likely due to the transient period necessary for the system to attain a steady state, coupled with the segregation of dissolved gas at the solidification front and the attainment of the critical nucleation concentration $C_n^*$. In the experiments involving increasing velocities, the gas segregation is pre-established from the onset when the velocity transitions from $18\mu m/s$ to $20\mu m/s$. Conversely, in the decreasing velocity ramps, it is probable that the steady state is not achieved for the initial velocity examined ($20\mu m/s$).

To a first approximation, we observed no influence of the front shape (planar or cellular) on the formation, growth, or engulfment of bubbles. With a cellular front, gas-enriched liquid is trapped between the cells, which might suggest a lower gas concentration at the front compared to a planar front. In principle, this could influence the cascade of bubble-related events at the front. However, since the front morphology is primarily determined by the front velocity, we cannot independently control both parameters and therefore cannot draw any conclusion on this point. Although a sharp transition from a planar to a cellular solidification front might be expected, we did not observe it here. For $V_{sf} < 5\mu m/s$ and $V_{sf} > 15\mu m/s$, the solidification front remains systematically planar and cellular, respectively. Instead of an abrupt change, a continuous evolution from planar to cellular solidification is observed, with both morphologies coexisting for intermediate velocities $8 < V_{sf} < 13\mu m/s$.

\subsection{Dynamics of bubble nucleation \label{ssec:dynamics_of_nucleation}}

\begin{figure}
  \centering
  \includegraphics[width=8cm]{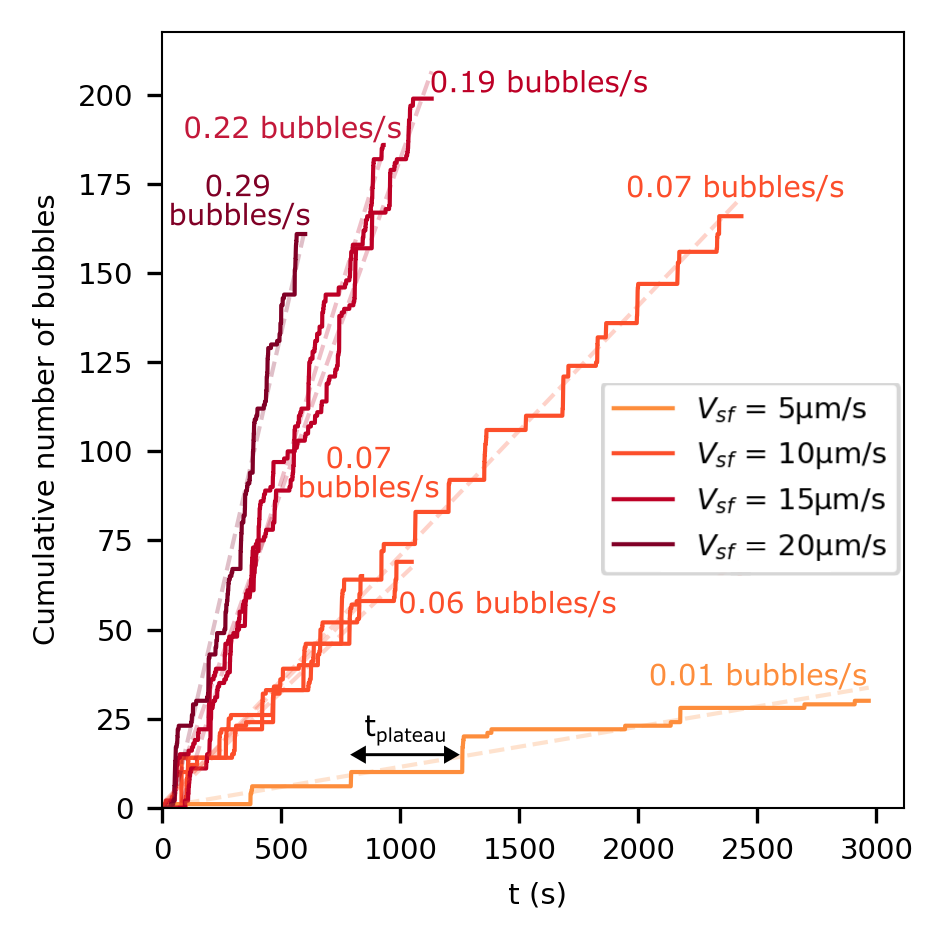}
  \caption{Cumulative number of nucleated bubbles as a function of solidification duration under varying velocities. Multiple experiments on different samples prepared following the same procedure were performed at solidification velocities of $V_{sf} = 10\mu m/s$ and $V_{sf} = 15\mu m/s$ to validate the reproducibility of the measurements. The plateau length indicates the lag period $t_{plateau}$ between successive bubble nucleation events.
}
  \label{fig:figure4}
\end{figure}

We now focus on the dynamics of bubble nucleation as influenced by $V_{sf}$ under experimental conditions where the solidification velocity $V_{sf}$ is maintained constant throughout the entirety of a given sample. In each captured image, the count of bubble nucleation events occurring within the field of view is recorded, and their cumulative total is evaluated over the duration of solidification for various solidification velocities (Fig.~\ref{fig:figure4}).

The slopes of these curves represent the average nucleation rate of gas bubbles for each specified solidification velocity $V_{sf}$, which concurs with the data presented in Fig.~\ref{fig:figure3}. Repeated experiments performed at solidification velocities of $10\mu m/s$ and $15\mu m/s$ consistently yielded convergent results, thereby verifying the reproducibility of the observed nucleation dynamics under controlled conditions. These findings confirm the reliability of the protocol employed for sample preparation and freezing. The rate of nucleation increases with the solidification velocity $V_{sf}$. For example, at $V_{sf}=5\mu m/s$, 30 bubbles nucleated after 2968s, in contrast to 166 bubbles after 2430s for $V_{sf}=10\mu m/s$, 186 bubbles after 930s for $V_{sf} = 15\mu m/s$, and 161 bubbles after 600s for $V_{sf}=20\mu m/s$.

Across the various solidification velocities investigated, the cumulative number of nucleated bubbles increases in a stepwise manner over time. The duration of the observed plateaus depends on the solidification velocity, with plateau durations noted as 380 $\pm$ 130s for $V_{sf}=5\mu m/s$, $131 \pm 43$s for $V_{sf}=10\mu m/s$, 58 $\pm$ 23s for $V_{sf} = 15\mu m/s$, and 40 $\pm$ 11s for $V_{sf}=20\mu m/s$.

This analysis reveals that the rate of bubble nucleation is non-uniform during the solidification process and occurs in discrete steps. These steps are governed by two alternating periods: (1) a lag period, $t_{plateau}$, in which gas accumulates at the solidification front, as indicated by the plateau observed in the nucleation curves, and (2) a burst nucleation phase, characterized by the simultaneous nucleation of numerous bubbles, corresponding to the steep increase in the curve. A reduction in gas segregation time preceding bubble nucleation is observed with an increase in the solidification velocity $V_{sf}$. This phenomenon arises because the segregation of dissolved gas at the solidification front becomes more significant at higher solidification velocities. Consequently, the magnitude of each nucleation event, as evidenced by the simultaneous nucleation of multiple bubbles, is also augmented.

It should be noted that the typical duration of plateau $t_{plateau}$ (tens to hundreds of seconds) is much longer than the time interval between successive frames (typically 0.13 s). We also systematically observe that as soon as bubbles appear, they go through a very fast growth stage and reach within a few frames. The stepwise characteristic of the dynamics observed in Fig.~\ref{fig:figure4} is therefore not due to spatial or temporal resolution issue (the appearance of the very small bubbles that would not be recorded at sufficiently shorter time scale). Hence, the steps reflect the real-time results of the bubble nucleation.

\begin{figure*}
  \centering
  \includegraphics[width=14cm]{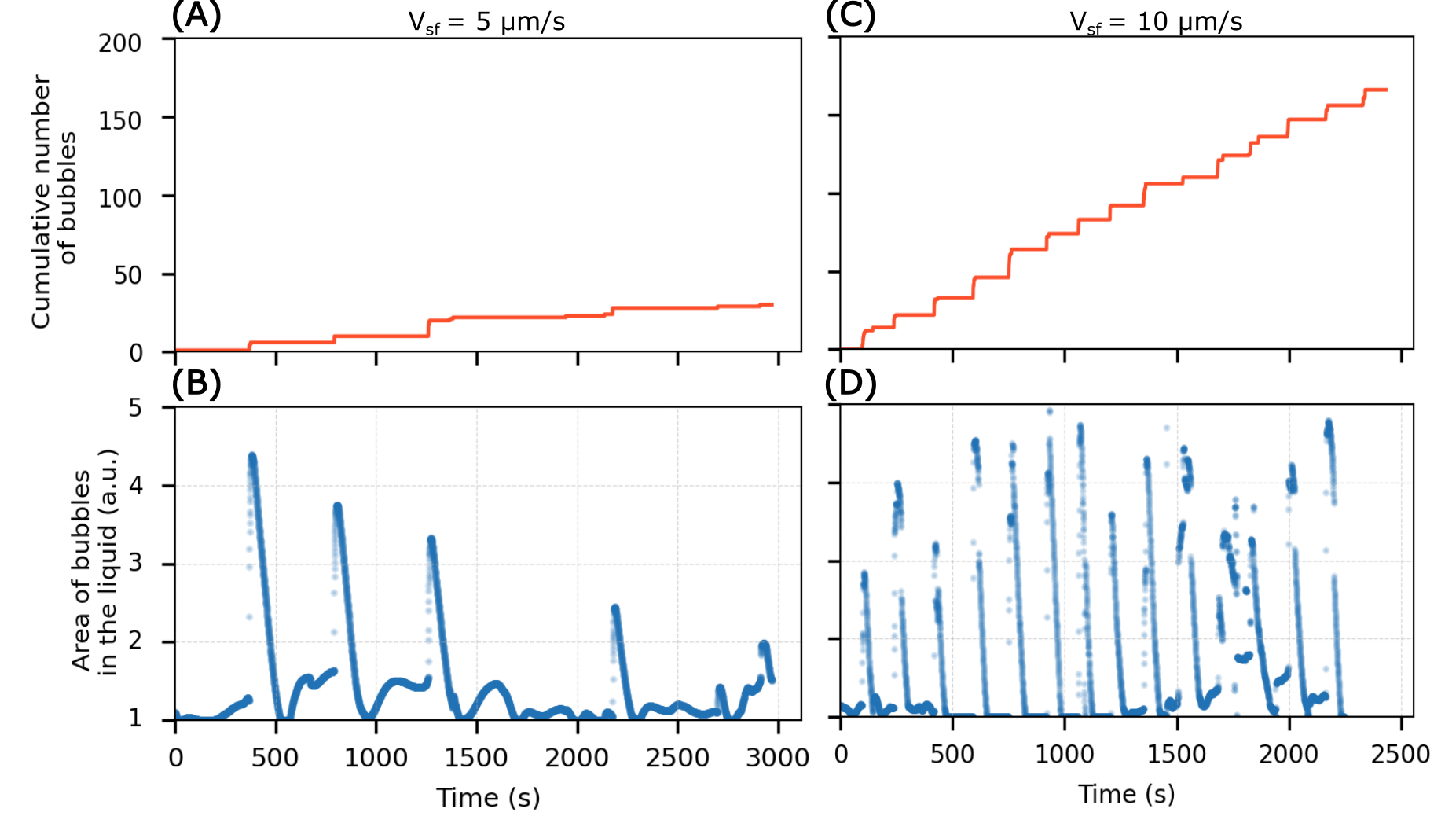}
  \caption{Evolution of the area occupied by bubbles and the sum of nucleated bubbles in the field of view of the microscope as a function of the duration of the solidification. (A), (C) Cumulative number of nucleated bubbles as a function of time (A) for $V_{sf}=5\mu m/s$ and (C) $V_{sf}=10\mu m/s$. (B), (D) Area occupied by nucleated bubble in the liquid as a function of the time for (B) $V_{sf}=5\mu m/s$ and (D) $V_{sf}=10\mu m/s$}
  \label{fig:figure5}
\end{figure*}

We now re-examine the data presented in Fig.~\ref{fig:figure4} to determine the total proportion of the field of view occupied by the bubbles in the liquid region. This analysis, depicted in Fig.~\ref{fig:figure5}, allows for the differentiation between the step of bubble nucleation and growth—resulting in an increase in the area occupied by the bubbles—and the subsequent step where they are engulfed by the front, leading to a reduction in the interfacial area. This method utilizes 2D fluorescence microscopy images, applying a threshold to identify and quantify the area occupied by the bubbles in the liquid region. The results, as a function of the duration of directional solidification corresponding to solidification velocities $V_{sf}=5\mu m/s$ and $10\mu m/s$, are presented in Fig.~\ref{fig:figure5}.

At all velocities, successive peaks of varying amplitudes are observed, characterized by rapid expansions of the space occupied by bubbles, followed by a slower decrease. Notably, the onset of the growth phase for each peak aligns with the increasing segments of the time evolution of bubble numbers as depicted in Fig.~\ref{fig:figure4}. These ascending sections of the peaks are attributed to the quasi-simultaneous nucleation of bubbles and their subsequent growth via gas diffusion from the supersaturated liquid. Conversely, the descending segments correspond to the gradual engulfment of these bubbles by the advancing solidification front, resulting in a reduction of their apparent area within the liquid. These curves are significant as they reflect the total area occupied by bubbles in the liquid. Thus, even when bubble growth is dominant, some bubbles may already start being engulfed by the solidification front. Conversely, when the area decreases, indicating that most bubbles are being engulfed, some bubbles may continue to grow simultaneously, even during their engulfment (as is the case for cylindrical bubbles, for example). The peak's maximum represents the transition point where the bubble growth rate is surpassed by the rate of their engulfment by the solidification front. The plateaus depicted in Fig.~\ref{fig:figure5} denote not only a period of gas segregation at the solidification front but also a phase during which the majority of the space occupied by the bubbles is engulfed by the advancing solidification front. The characteristic lag time, $t_{plateau}$ (Fig.~\ref{fig:figure4}), can thus be interpreted as the cumulative duration of bubble nucleation and growth $t_n$ (ascending part of the peaks), the bubble engulfment period $t_e$ (descending part of the peaks), and a subsequent interval of gas segregation without significant nucleation or engulfment activity.

Furthermore, the minor oscillations observed between the main peaks (Fig.~\ref{fig:figure5}B) are primarily ascribed to the existence of cylindrical bubbles at the solidification front. In these instances, a portion of the bubble remains emerged in the liquid, sustained by the supply of dissolved gas. These cylindrical bubbles may achieve a quasi-equilibrium state with a consistent width, although the diameter of the emerged tip can fluctuate based on pressure, gas diffusion, and the velocity of solidification. Consequently, in the latter scenario, the area occupied by the emerged tip of the bubble evolves over time. As illustrated in Fig.~\ref{fig:figure2}B and Fig.~\ref{fig:figure2}C, cylindrical bubbles persist longer at the interface and display larger diameters at reduced solidification velocities. Under these circumstances, their extended presence at the interface leads to a greater consumption of gas, as the diffusion rate of gas sufficiently compensates for the engulfment by the advancing solidification front. This gas diffusion towards cylindrical bubbles may consequently decelerate the rise in gas concentration at the solidification front. Thus, we assume that the presence of cylindrical bubbles influences bubble nucleation dynamics, particularly affecting the bubble nucleation rate during solidification. Nonetheless, the nucleation of a solitary bubble can also occur, contributing to these minor oscillations between the peaks.

\begin{figure*}
  \centering
  \includegraphics[width=16cm]{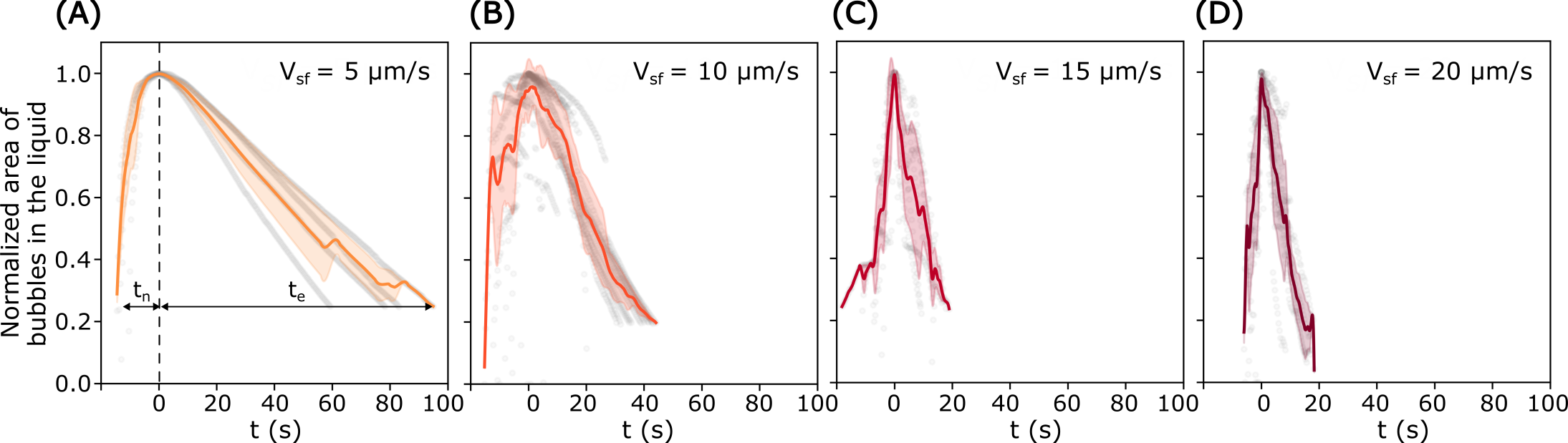}
  \caption{Temporal evolution of the area occupied by bubbles during the stages of nucleation, growth, and engulfment throughout the solidification process. The peaks identified in the global measurements of the bubble-occupied area during solidification were extracted and normalized relative to their maximum values. The boundaries for each peak were determined as the times at which the measured area decreased to below 20\% of its maximum. Different colors in the graphs represent distinct peaks, each corresponding to a specific cycle of bubble nucleation, growth, and subsequent engulfment. (A) $V_{sf}=5\mu m/s$ (B) $V_{sf}=10\mu m/s$ (C) $V_{sf}=15\mu m/s$ (D) $V_{sf}=20\mu m/s$. The number of peaks extracted is contingent on the solidification velocity, given that the characteristic nucleation time diminishes as solidification velocity increases. Consequently, 4 peaks were extracted for 2968s with $V_{sf}=5\mu m/s$, 13 peaks for 2430s with $V_{sf}=10\mu m/s$, 9 peaks for 930s with $V_{sf}=15\mu m/s$, and 10 peaks for 600s with $V_{sf}=20\mu m/s$.}
  \label{fig:figure6}
\end{figure*}

\begin{figure}
  \centering
  \includegraphics[width=8cm]{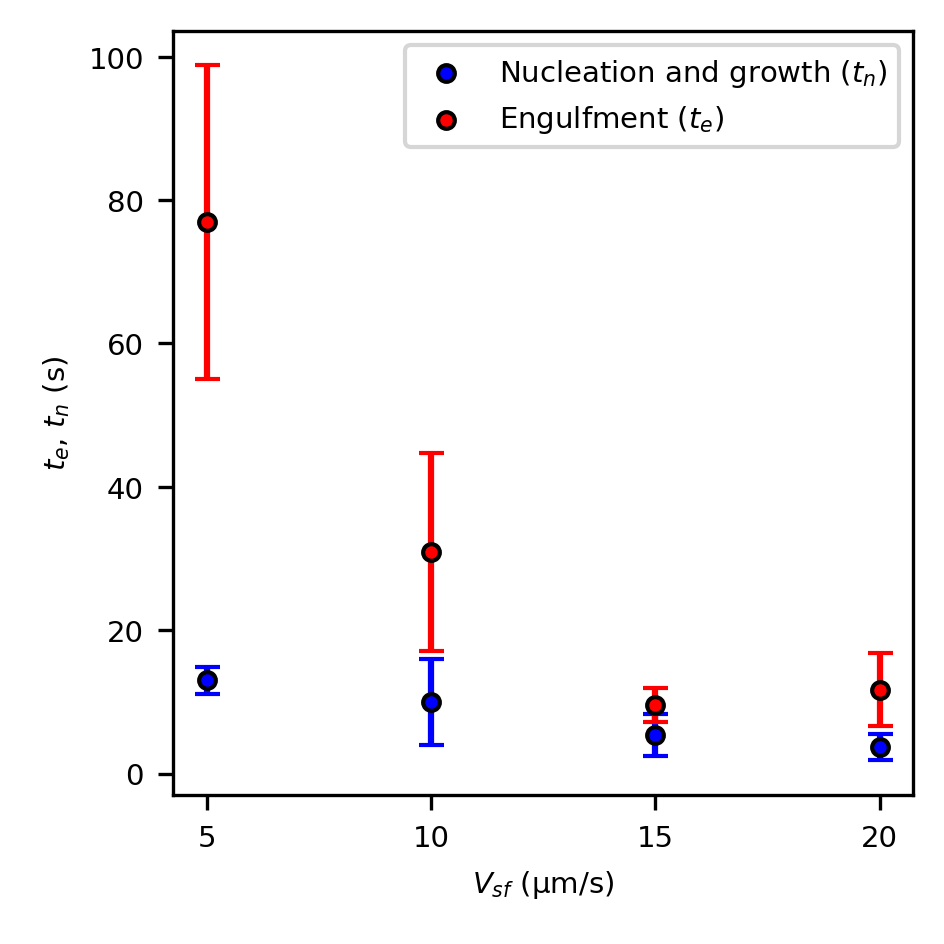}
  \caption{Temporal dynamics of gas bubble evolution during solidification as a function of solidification velocity $V_{sf}$. Blue points represent time intervals during which bubble nucleation and growth are preeminent compared to the engulfment action of the solidification front. Conversely, red points signify intervals during which the engulfment of bubbles dominates.}
  \label{fig:figure7}
\end{figure}

Based on Fig.~\ref{fig:figure5}, we extracted the peaks describing the evolution of the area occupied by the bubbles and normalized them (Fig.~\ref{fig:figure6}). This methodology enables an accurate comparison of bubble dynamic profiles and quantifies how variations in solidification velocity influence the overall bubble behavior throughout the solidification process. Our findings reveal that for all investigated solidification rates, the observed peaks and corresponding durations of nucleation, growth, and engulfment periods of bubbles remain consistent for a given solidification velocity. For solidification velocities of $V_{sf}=15\mu m/s$ and $V_{sf}=20\mu m/s$, a notable symmetry emerges, indicating that the duration of bubble nucleation and growth $t_n$ matches that of bubble engulfment $t_e$. At lower solidification velocities, particularly for $V_{sf}=5\mu m/s$ and $V_{sf}=10\mu m/s$, the peaks exhibit asymmetry due to an extended duration of the bubble engulfment phase as the solidification velocity decreases. The analysis of these peaks facilitates the determination of the timescales linked with the coupled gas and bubble dynamics, which are influenced by the solidification front velocity (Fig.~\ref{fig:figure7}).

We observe a reduction in the average duration of gas bubble nucleation and growth from $t_n=13.s$ at $V_{sf}=5\mu m/s$ to $t_n=3.5s$ at $V_{sf}=20\mu m/s$. Furthermore, a pronounced decline is noted in the duration of bubble engulfment by the solidification front, decreasing from $t_e=76.9s$ at $V_{sf}=5\mu m/s$ to $t_e=11.7s$ at $V_{sf}=20\mu m/s$.

This indicates that the duration of bubble nucleation and growth is less influenced by the solidification velocity compared to the time necessary for their engulfment by the solidification front. At elevated solidification velocities, bubbles are more swiftly engulfed due to the faster progression of the solid--liquid interface. Furthermore, when the solidification velocity attains a sufficiently high magnitude, the diffusion of gas from the surrounding liquid is inadequate to compensate for the effect of the advancing front, thus imposing a limitation on bubble growth. Conversely, at reduced velocities, bubbles can grow for extended periods as the interface progresses at a slower rate. As will be elaborated in the subsequent section, the fact that the duration of bubble nucleation and growth is less affected by the solidification velocity is attributable to the observation that the critical gas concentration required for nucleation remains relatively constant within our system.

The impact of the coalescence of adjacent bubbles should also be briefly discussed. It is well-known that there is a critical size (for a given front velocity) below which bubbles are repelled by the front, and above with they engulfed by the front~\cite{Park2006}. One could therefore imagine a situation where small bubbles (below the critical size) would coalesce and therefore be engulfed by the front. In such situation, bubbles coalescence could impact the results reported here. However, we systematically observed rapid growth of bubbles and quasi-systematic engulfment of the bubbles by the front. In other words, bubbles here are almost never small enough to be repelled by the front. We are therefore never in a scenario where the coalescence of bubbles would impact their interaction with the solidification front.

We focused so far on the nucleation rate and characteristic nucleation times as a function of the solidification velocity. Nevertheless, an additional crucial parameter warranting investigation is the spatial location of bubble nucleation.

\subsection{Where do bubbles nucleate ?\label{ssec:nucleation_location}}

We direct our focus toward the location of bubble nucleation during the solidification process. The nucleation distance was quantified as the distance between the bubble and the solidification front in the first frame where the bubble appeared visible. We postulate that bubble movement immediately following nucleation and between frames is negligible. This assumption is corroborated by our experimental findings, as no movement of bubbles was observed between nucleation and their interaction with the solidification front or adjacent bubbles. 

\begin{figure*}
  \centering
  \includegraphics[width=16cm]{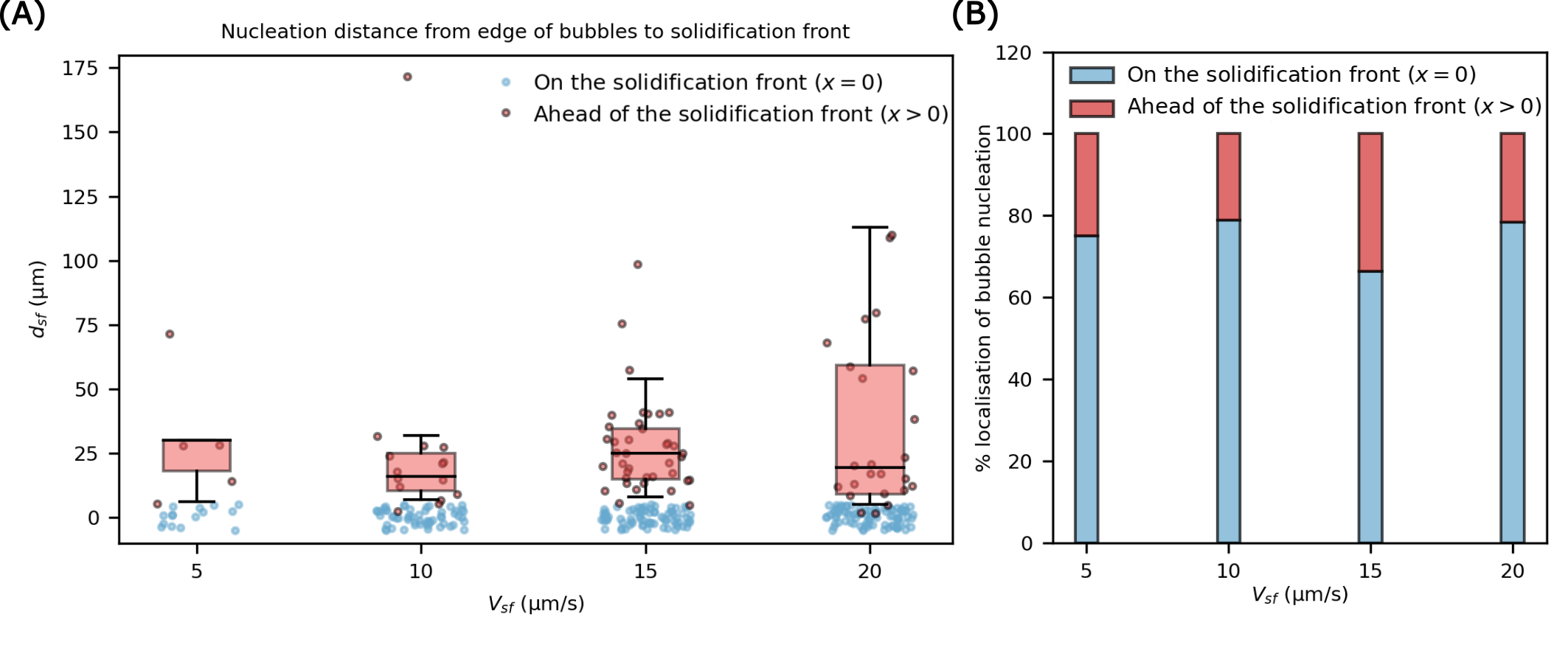}
  \caption{Bubble nucleation distance from the solidification front. (A) Measurement of bubble nucleation distances $d_{sf}$ for different solidification velocities $V_{sf}$. These distances are estimated from the initial appearance of the bubble, with the assumption that the displacement of the bubble from its nucleation point to its first observable position is negligible. Measurements were conducted for four different applied solidification velocities, and lateral dispersion was integrated into the data points to improve visual clarity. (B) Proportion of bubbles that nucleate at the solidification front and ahead of the front in relation to solidification velocity $V_{sf}$.}
  \label{fig:figure8}
\end{figure*}

The distance between the solidification front and either the center or edge of the bubble was quantified in relation to the solidification rate $V_{sf}$ (Fig.~\ref{fig:figure8}). Our findings reveal that approximately 73\% of the bubbles nucleated during the experiments emerged on the solidification front. This observation suggests a predominantly heterogeneous nucleation mechanism, facilitated by the presence of a substrate (the ice crystals) and the locally elevated gas concentration close to the solidification front due to gas segregation. 

It is frequently asserted in solidification studies that all gas bubbles nucleate heterogeneously either on the solidification front~\cite{Maeno1967-ck,Wei2004-xa} or on particles present within the liquid~\cite{Werner2021-gc,Kato1999nucleation}. Moreover, theoretical analyses propose that nucleation is considerably more probable at the solidification front due to its favorable energetic conditions~\cite{Murakami2002-al,Bunoiu2010-hw}. Our results are therefore in good agreement with previous studies.

We found that 27\% of the bubbles nucleate within the liquid phase, predominantly within the region extending up to $100~\mu m$ from the solidification front, where gas segregation initiates. The solidification velocity does not affect this proportion, as the percentage of bubbles nucleating on the solidification front consistently remains between 60\% and 80\% across all tested velocities, without any discernible trend (Fig.~\ref{fig:figure8}B). Some empirical observations suggest that while the predominant majority of bubbles nucleate heterogeneously on the front, a minor portion may nucleate in proximity to it without the front serving as a substrate~\cite{meijer2024bubble}. Such nucleation events are typically interpreted as heterogeneous nucleation occurring on minor impurities~\cite{ZHDANOV1980659}.

\subsection{Critical gas concentration for nucleation\label{ssec:nucleation_concentration}}

\begin{figure*}
  \centering
  \includegraphics[width=14cm]{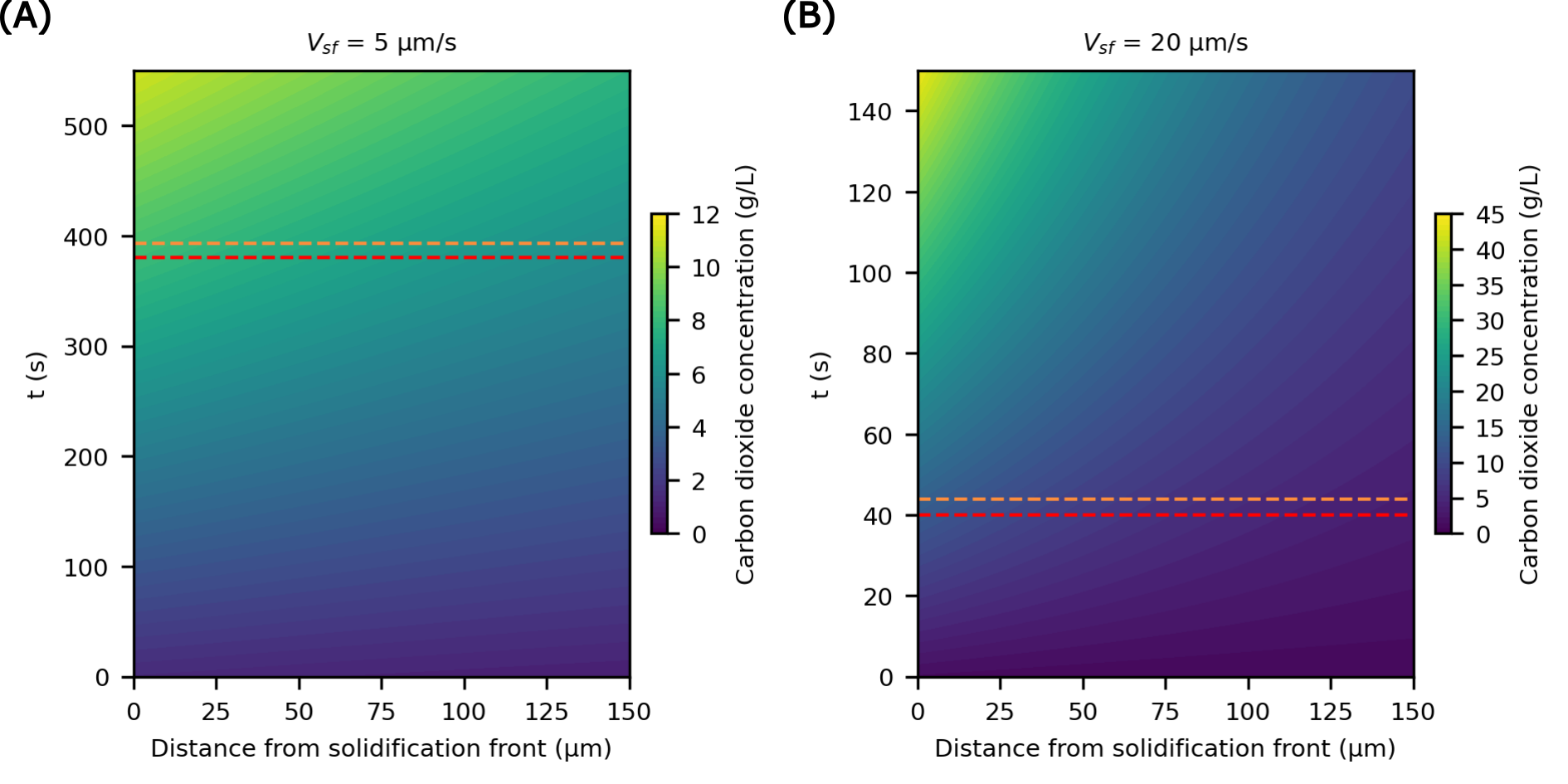}
  \caption{Gas segregation profile at the solidification front during solidification process and estimation of the critical nucleation concentration in our system. The segregation profiles were derived from equation~\ref{eq:concentration_profile}, specifically for the instance of carbon dioxide dissolved in water at an initial concentration $C_0 = 1.4g/L$. The critical concentration estimates were obtained using the average nucleation distances and the characteristic nucleation times for the red dashed lines, in addition to the sum of the characteristic nucleation time $t_n$ and the nucleation time for the orange dashed lines. (A) $V_{sf} = 5\mu m/s$ (B) and $V_{sf} = 20\mu m/s$.}
  \label{fig:figure9}
\end{figure*}

The critical bubble nucleation concentration $C_n^*$ within our system can be derived from experimental observations, using the bubble nucleation distance and the characteristic nucleation time under varying solidification conditions. Assuming that each nucleation event resets the local conditions and initiates a novel phase of gas segregation at the solidification front, Pohl's equation (Eq.~\ref{eq:concentration_profile}) is used to estimate the gas concentration at the specific time and distance to the solidification front (Fig.~\ref{fig:figure9}). The interval required to achieve the critical gas concentration facilitating further nucleation corresponds to the lag time $t_{plateau}$ depicted in Fig.~\ref{fig:figure4} and Fig.~\ref{fig:figure5}. To accurately assess the critical concentration for bubble nucleation, it is crucial to ascertain the nucleation site of the bubbles to incorporate the corresponding distance from the solidification front, denoted as x, into equation~\ref{eq:concentration_profile}. 

Under our experimental conditions, 73\% of the bubble population nucleates heterogeneously at the solidification front ($x = 0$). We therefore focus first on this population of bubbles. Under these conditions (Tab.~\ref{tab:nucleation_concentration}), the critical concentration estimated using Eqn.~\ref{eq:concentration_profile} is found to range from  $C_n^* = 8.2 \pm 0.1~g/L$ for $V_{sf}=5~\mu m/s$ to $C_n^* = 13.3 \pm 0.7~g/L$ for $V_{sf}=20~\mu m/s$.

Following nucleation, the majority of these bubbles expand prior to being eventually engulfed by the front. As a result, a significant fraction of the gas that segregates at the solidification front is consumed through both the heterogeneous nucleation of bubbles and their subsequent growth, regardless of the nucleation pathway involved. These findings indicate that the effective gas concentration at the interface preceding bubble nucleation may decrease below the level predicted by equation~\ref{eq:concentration_profile}, suggesting that the true critical bubble nucleation concentration $C_n^*$ under our experimental conditions might be lower than currently estimated.

\begin{table*}
\begin{ruledtabular}
    \centering
    \begin{tabular}{ccccc}
         $V_{sf}$  & $t_{plateau}$  & $t_{plateau}$  &  nucleation distance x  & $C_n^*$ (g/L) \\
        ($\mu m/s$)  &(s) & + nucleation time (s) & to the front ($\mu m$)&  \\ 
          \hline
        5 & $380 \pm 130$  & $ 393\pm 132 $ &  0  & $8.2 \pm 0.1$\\
        10 & $ 131\pm 43$ & $ 141\pm 53$ &  0 & $11.1 \pm 0.4$\\
        15 & $ 58\pm 23$ & $ 64\pm 26$ &  0 & $11.2 \pm 0.6$\\
        20 & $ 40\pm 11$ & $ 44\pm 13$ &  0 & $13.3 \pm 0.7$\\
    \end{tabular}
    \caption{Estimation of the critical concentration for bubble nucleation for bubbles nucleating on the front. This table provides the values of the characteristic nucleation time between successive bubble nucleation bursts $t_{plateau}$, the nucleation time, and the nucleation distance of bubbles from the solidification front (x=0 when nucleation occurs on the front). The critical concentrations for bubble nucleation $C_n^*$ are estimated using equation~\ref{eq:concentration_profile}, using the experimentally obtained values presented in this table.}
    \label{tab:nucleation_concentration}
    \end{ruledtabular}
\end{table*}

We focus now on the bubbles nucleating ahead of the solidification front. Among the range of solidification velocities explored, the principal characteristics distinguishing the critical nucleation concentration are the gas segregation profile and the characteristic lag time t\textsubscript{plateau}. The estimation are provided in Tab.~\ref{tab:nucleation_concentration_ahead}. Our findings indicate that the critical nucleation concentration for bubble formation $C_n^*$ remains consistent across the three solidification velocities examined, demonstrating nearly identical $C_n^*$ values for $V_{sf}=10~\mu m/s$, $V_{sf}=15~\mu m/s$, and $V_{sf}=20~\mu m/s$. The estimated critical nucleation concentration for $V_{sf}=5~\mu m/s$ deviates slightly from those obtained at alternative solidification velocities, with a value approaching $4.5~g/L$. Given that the nucleation distance seems to remain largely unaffected by changes in solidification velocity, this variance might originate from the distinctive lag time $t_{plateau}$ observed at this particular velocity. Importantly, $V_{sf}=5~\mu m/s$ denotes the scenario where nucleation events were the least frequent, and bubble growth at the interface was notably pronounced, thereby resulting in limited statistical data. Hence, the critical nucleation concentration estimated under these circumstances may be less reliable than at other velocities. Nevertheless, the estimated value retains the same order of magnitude as those derived from other solidification velocities, which we consider satisfactory given the extensive range of critical nucleation concentrations reported in the literature.

Therefore, for the other three velocities examined, it is inferred that the rate of solidification has an insignificant impact on the critical concentration required for bubble nucleation. Analysis of the entire set of experimental data leads to the estimation of an average value $C_n^* = 8.4 \pm 3.1~g/L$, which exceeds the carbon dioxide solubility limit in water at 25~$^{\circ}$C by approximately 5 to 6 times, and approaches 3 times the solubility limit at 0~$^{\circ}$C.

\begin{table*}
\begin{ruledtabular}
    \centering
    \begin{tabular}{cccccc}
         $V_{sf}$  & $t_{plateau}$  & $t_{plateau}$  & nucleation distance  & nucleation distance  & $C_n^*$ (g/L) \\
        ($\mu m/s$)  &(s) & + nucleation time (s) & (center) ($\mu m$)& (bottom) ($\mu m$)& \\ 
          \hline
        5 & $380 \pm 130$  & $ 393\pm 132 $ & $ 67\pm 41 $ & $ 26\pm 21$ & $ 4.5\pm 2.1$\\
        10 & $ 131\pm 43$ & $ 141\pm 53$ & $ 78\pm 50$ & $ 24\pm 33$ & $ 8.7\pm 3.2$\\
        15 & $ 58\pm 23$ & $ 64\pm 26$ & $ 63\pm 23$ & $ 27\pm 18$ & $ 8.1\pm 3.0$\\
        20 & $ 40\pm 11$ & $ 44\pm 13$ & $ 72\pm 29$ & $ 31\pm 29$ & $ 8.2\pm 3.1$\\
    \end{tabular}
    \caption{Estimation of the critical concentration for bubble nucleation for bubbles nucleating ahead of the front. Experimentally measured characteristic data on bubble nucleation dynamics in our system. This table provides the values of the characteristic nucleation time between successive bubble nucleation bursts $t_{plateau}$, the nucleation time, and the nucleation distance of bubbles from the solidification front, with measurements taken at the middle and bottom of the nucleated bubbles. The critical concentrations for bubble nucleation $C_n^*$ are estimated using equation~\ref{eq:concentration_profile}, using the experimentally obtained values presented in this table.}
    \label{tab:nucleation_concentration_ahead}
    \end{ruledtabular}
\end{table*}

Regardless of the location of bubble nucleation, on the front or ahead of the front, our results therefore provide a similar estimate of the critical gas concentration for nucleation. The measurement of the critical nucleation concentration in directional solidification presents notable challenges. Various estimates suggest it ranges from 7~\cite{geguzin1981crystallization} to 20~\cite{lipp1987gases}, 40~\cite{Thievenaz2025bubble} or 100 times~\cite{jones1999bubble} the solubility limit, depending on the solidification conditions. Our findings ($\approx$ 3 times the solubility limit), therefore, fall within the lower spectrum of the estimated critical nucleation concentrations for directional solidification. The phenomenon of bubble nucleation predominantly occurring in bursts suggests that the initial bubble in a sequence may perturb the surrounding liquid, thereby inducing either a localized supersaturation or mechanical perturbations conducive to subsequent nearby nucleation. Although this  hypothesis remains plausible, we were unable to confirm it experimentally. 

\section{Conclusions\label{sec:conclusions}}

In summary, cryo-confocal fluorescence microscopy was used to conduct an \textit{in situ} investigation of the coupled dynamics of gas segregation and bubble nucleation, growth and engulfment during directional solidification under rigorously controlled conditions, focusing particularly on solidification velocities and temperature gradients. We demonstrated that the dynamics of bubble nucleation are governed by a characteristic nucleation time, which is determined by the solidification velocity. This characteristic time emerges from the interaction of several mechanisms occurring during solidification, including gas segregation at the solidification front, bubble nucleation and growth, and their subsequent engulfment by the front. The parameters of solidification further influence the equilibrium between gas segregation and the engulfment rate, which in turn affects the interaction of bubbles with the solidification front, thereby impacting the overall bubble nucleation dynamics.

Nucleation was found to be predominantly heterogeneous, occurring on the solidification front. These experimental observations enabled us to propose an estimation for the critical gas concentration for bubble nucleation within our system— a critical value that appears to be unaffected by the solidification velocity. These findings enhance our comprehension of the fundamental processes governing bubble nucleation dynamics during solidification, thus contributing to the improved control over porosity formation in solidified materials.

\begin{acknowledgments}
We acknowledge the CNRS for its financial support of the COBLE project through the PRIME80 program, and the ANR - FRANCE (French National Research Agency) for its financial support of the FROST project ANR-22-CE06-0022.
\end{acknowledgments}

\textbf{CRediT author statement: }
\textbf{Bastien Isabella}: Investigation, Validation, Formal analysis, Writing - Original Draft, Visualization
\textbf{Cécile Monteux}: Conceptualization, Writing - Review \& Editing, Supervision, Funding acquisition
\textbf{Sylvain Deville}: Conceptualization, Writing - Review \& Editing, Supervision, Funding acquisition

\bibliography{biblio.bib}

\end{document}